# Soil Stabilization and Renewal Using Annealed Polyvinyl Alcohol as Long-lasting Binder

Chunyan Cao[1]* and Gang Li[2]*

**Affiliations:**

[1] School of Electrics and Computer Engineering, Nanfang College, Guangzhou, Guangzhou 510970, China

[2] Department of Materials Science and Engineering, Southern University of Science and Technology, Shenzhen 518055, China

*Corresponding author. Email: caochy@nfu.edu.cn and 11930690@mail.sustech.edu.cn

**Abstract:** Soil, a crucial yet non-renewable resource for agriculture and the environment, is rapidly eroding due to human activities. To address this issue, we enhance the stability of the soil with polyvinyl alcohol (PVA) through a mixing, drying, and annealing process. The PVA-treated soil can resist water impact at 7 m/s, preventing rainfall erosion. The water-retaining capability and drainage of PVA-treated soil can be adjusted by changing the size. No deterioration is observed after eight months of cultivation. With this approach, we can mitigate soil erosion, customize soil properties, and renew eroded soil, offering a strategy to reverse the trend of soil depletion.

**One-Sentence Summary:** Skin Care and Regeneration for Mother Earth

## Main Text:

Soil is the skin of the Earth, the foundation of crop production, and a vital resource for human society. (*1-5*) Agricultural practices eliminate ground cover, leaving the exposed soil vulnerable to rainfall-induced erosion.(*6-10*) The Food and Agriculture Organization of the United Nations (FAO) 2015 report estimates that 20-30 gigatons of soil are eroded per year.(*11*) In tropical and subtropical hilly croplands, the eroded soil accounts for up to 50-100 t ha$^{-1}$ yr$^{-1}$, which is 100 times faster than soil formation.(*6, 11-13*) Conservation practices are proposed to mitigate soil erosion, with an estimated reduction of about 7.1%.(*13*) However, conservation practices not only lack efficiency in tackling soil erosion but are also unaffordable in less developed countries. The imbalance between rapid erosion and slow production will ultimately lead to the final depletion of soil, especially in less wealthy areas. Existing soil protection measures are insufficient to reverse the ongoing trend of soil depletion as long as we still depend on nature for soil production. We need to reduce erosion more efficiently and produce soil on our own to achieve soil sustainability.

We optimize the natural soil formation process and find a better soil binder that can stabilize soil and renew eroded soil. The key step in soil formation is the construction of soil aggregate, where nature utilizes cementing agents, such as humus, to bind soil particles into aggregates.(*14*) However, the weak adhesion and biodegradable nature of humus fail to sustain the integrity of exposed soil under the impact of raindrop.(*15, 16*) Instead of humus, we apply a synthetic polymer polyvinyl alcohol (PVA) to bind soil through mixing, drying, and annealing. The annealed PVA strongly binds to the clay particles in soil,(*17-19*) which creates a high cohesive strength in PVA-soil. The PVA-soil can withstand the impact of water at a speed of 7 m/s, keeping it from water induced decomposition. The high stability in water of PVA-soil leads to adjustable physical properties. For instance, the water-retaining capacity (in terms of wetness) of PVA-soil can be adjusted from 27 wt.% to 70 wt.% by changing the size from 5-10 mm to 0.5-2 mm, along with a two-magnitude shift of hydraulic conductivity from 4.5 cm/s to 0.26 cm/s. In addition, our method works in eroded soil such as sediment and capable to endow the eroded soil with preferrable water retaining capacity and drainage, making them useful planting substrate. In brief, our method can reduce soil erosion, define soil property, and make eroded soil renewable, which opens a new avenue for soil conservation and management.

**Stability in water**

We treat soil (fig. s1) by mixing a PVA (KURARAY POVAL™ 60-98, 98-99 mol% hydrolysis) solution with soil (2 wt.% PVA in soil), followed by drying and annealing. The mixing, drying and annealing treatments improve the stability of PVA (fig. s2) and result strong interaction between PVA and clay particle (fig. s3). After all, a PVA-soil is obtained. The PVA-soil retains its grain shape without changes after soaking in water, whereas the untreated soil crumbles into a mud-like consistency (fig. s4). The PVA-soil, once wetted, remains unchanged after drying, while the untreated soil forms a crust and develops cracks (fig. s5). In the wet sieving test, the untreated soil breaks down and passes through the mesh. In the contrast, PVA-soil has no mass loss occurs (fig. s6).

We examine the stability of PVA-soil in extreme condition, such as high-speed water impact (Movie S1) and high-frequency shaking (Movie S2). The PVA-soil grain under a 7 m/s water impact shows no fracturing (Fig. 1a) and remains its physical integrity after a 5-minute test (Fig. 1b). Quantitatively, the PVA-soil lost <0.2% and 0.8% mass at a water speed of 1 m/s and 3 m/s, respectively. And when the speed rises to 7 m/s, the mass of detached

PVA-soil is 3%. The photos of PVA-soil before and after 1 hour of shaking at 4 Hz in Figure 1c and d show only minor disintegration of PVA-soil from the grains. At low shaking frequencies of 1.33 Hz and 2.67 Hz, the reduction in mass of PVA-soil grain is less than 0.5% (Fig. 1f). At a high shaking frequency of 4 Hz, an average of 1.7% of PVA-soil is broken into the water (Fig. 1f). In comparison, most natural soils disintegrate during wet sieving, not to mention undergoing rigorous flushing and shaking tests.

**Customize soil property**

We evaluate the porous structure of PVA-soil with different sizes. We crush and sieve the PVA-soil into three classes, namely, PVA-soil-S (0.5-2 mm), PVA-soil-M (2-5 mm) and PVA-soil-L (5-10 mm) (Fig. 2a). The dimensions of the pores between the aggregates in these three samples are estimated to be in the orders of sub-millimeter, millimeter, and sub-centimeter, respectively. In addition, the SEM image reveals internal pores of PVA-soil has large one of $10^2$ μm and small one of several μm (Fig. 2b). The porous characteristic reduces the density of PVA-soil from 2.5 g/cm$^3$ in particle form to 0.75 g/cm$^3$ in bulk (Fig. 2c). The PVA-soils exhibit a similar porosity of 70%, but the variation in inter-aggregate pores contributes significantly to other properties.

We test the water retaining capability and air fraction of the PVA-soils in saturated situation. A grain of PVA-soil can absorb 24.5 wt.% water (Fig. 2a). For PVA-soil-L, the wetness is 27.6 wt.%, only 3.1 wt.% higher than that of a single grain. In the case of PVA-soil-M and PVA-soil-S, the wetness is 39.5 wt.% and 76.2 wt.%, respectively. Their smaller inter-pores have retained more water. Pores can be filled with water or gas, and thus the differing wetness leads to a distinct solid-liquid-gas composition in the soil (Fig. 2e). The gas volume fraction in PVA-soil-L is 49.3 vol.%, while this fraction reduces to 13.0 vol% in PVA-soil-S (Fig. 2e). The pores between aggregates can be regulated by the size of aggregate, thus providing approach to adjusting small pores for water retaining or large pore for gas restore.

We study the drainage of PVA-soils by measuring their hydraulic conductivity in saturated state. The untreated soil is poor drainage, characterized by a low hydraulic conductivity of $6\times10^{-4}$ cm s$^{-1}$ (Fig. 3f). In contrast, PVA-soil-SS, featuring an aggregate size less than 0.5 mm, demonstrates a significantly higher conductivity of 0.014 cm/s. Even higher conductivities are observed in PVA-soil-S and PVA-soil-M, with hydraulic conductivities of 0.27 cm/s and 2.64 cm/s, respectively. The coarser sample, PVA-soil-L, attains the highest conductivity of 4.50 cm/s. In the untreated soil, the broken aggregate clog the pores and hinder water flow, resulting a poor drainage. Conversely, the water stable aggregates in PVA-soil hold pores in saturated state that facilitate water movement (fig. s7 and Movie s3). And different pores dimension determined by the different aggregate size thus show adjustable drainage.

We test the tolerant of our method with regard to mineral additive, such as vermiculite and perlite (figure s8). The addition of 30 vol.% vermiculite or perlite into PVA-soil do not deteriorate the impact resistance or drainage, yet it can reduce the bulk density and increase the wetness (Table 1). These porous minerals increase the porosity of soils, enabling them to retain more water than PVA-soil and provide higher volume fraction for air simultaneously (Table 1).

We extend the PVA treatment to more soils, like mud and red soil (fig. s9). PVA-mud and PVA-red soil prepared using the same route is showed in figure s10-15 and their properties is listed in Table 1. Both of the PVA-mud and PVA-red soil can withstand the water impact test, showing a minor mass loss of 4-6 wt.%. The porosity and wetness of these soils are closed to that of PVA-soil, and all exhibit good drainage. Because the basic mechanism of our method is to stabilize soil using PVA as clay binder, therefore it works well on other mineral soils.

**Long-term stability**

We assess the effectiveness and long-term stability of the soil through a planting experiment. Conducted in Chaozhou (116.6°E, 23.6°N), a subtropical city in China, the potted experiments involved different plant species with durations ranging from 2 to 8 months. Currently, all plants are exhibiting robust growth (figure s16). Taking the example of a potted jasmine that has been cultivated for eight months, the surface soils maintain their grain-shape over the extended period, accompanied by the growth of moss or lichen (Fig. 3a). The internal soil remains loose with visible pore spaces (Fig. 3b) and still possesses high permeability, and there is no evidence of compaction (Movie S4). When the plant roots are separated from the soil, the roots are dense and evenly distributed (Fig. 3c), indicating favorable conditions for both water retention and aeration, promoting good growth conditions for the plant roots.

We test the physical properties of the soil after eight months of cultivation. In the water impact resistance test, the surface aggregate exhibits minimal mass loss of <2 wt.% (Fig. 3d), possibly due to the protection of lichen or moss against water impact. The internal soil also showed no significant reduction in erosion resistance. Regarding wetness and hydraulic conductivity, the performance of the internal soil did not differ significantly from freshly prepared samples (Fig. 3e).

**Biodegredability and safety**

We assessed the biodegradability and eco-compatibility of PVA. Previous studies indicate that PVA is poorly degradable in soil,(*20-24*) primarily due to the scarcity of PVA-degrading microorganisms. In one study, PVA sheets were buried in 18 natural soil sites with varying compositions and climates for two years.(*23*) Their results revealed that the degraded PVA was less than 10%. This poor degradability ensures the long-term usage of PVA in soil. The second reason for poor biodegradability is attributed to the irreversible absorption of PVA on clay.(*21*) Another research shows that the interaction between PVA and clay can hinder the attack of organisms on PVA.(*18*) Only 4% of clay-absorbed PVA degraded in a solution containing PVA-degrading organisms after one month of incubation.(*18*) In the contrast, 34% of the free PVA was degraded.(*18*) This result implies that our PVA-soil can sustain for a long period even with PVA-degrading microorganisms present. Additionally, PVA is non-toxic and does not harm plants.(*25*) The high stability in soil and non-toxic nature of PVA ensure our method to be applied in practice with high safety and long-term stability.

**Mitigate soil erosion**

The PVA treated soil resists erosion in two ways. The PVA-treated soil withstands water impact at 7 m/s for 5 minutes and sustains 4 Hz reciprocal shaking in water for 1 hour (Fig. 1, Moive S1-2). Such high stability can prevent the decomposition by raindrops impact, inhibiting the initial stage of water erosion. Even if the soil is decomposed by other forces, such as machinery tillage, the broken PVA-treated soil maintains high drainage (Fig. 2f and fig. s7), facilitating water infiltration and delaying the formation of surface runoff—the second stage of water erosion that carries soil away.

In practice, our method is flexible. First, the dosage of PVA can be optimized based on soil texture and climatic conditions to achieve a balance between erosion resistance and economic cost. Second, the energy-consuming step, annealing treatment, can be replaced by air-drying. Eliminating residual water in the PVA-soil mixture in a dry environment also leads to high stability for PVA-soil.

**Customize soil properties**

Due to their inherent instability, natural soils cannot maintain consistent physical properties across varying water content. (fig. s4-5 and fig. s10-13). The optimization of soil properties for cultivation often relies on empirical methods rather than repeatable ones. Modern precision agriculture has achieved precise control over fertilizer, irrigation, and so on. However, the lack of control over soil properties limits further increases in productivity.

PVA treatment makes soil properties customizable. The solid phase in PVA-treated soil remains an unchanged skeleton during wet-dry variations (fig. s4-5 and fig. s10-13), allowing the regulation of physical properties by manipulating aggregate sizes (Fig. 2). This adjustable property enables control over soil wetness and drainage in a repeatable manner. We can prepare soil based on the requirements of different plants. For instance, we can create fine soils for plants with high water demand and produce coarse soils for those requiring significant aeration, such as orchids. In the past, we needed to select suitable plants for each type of soil; now, we are moving towards designing customized soils for each plant.

**Renew eroded soil**

Our method can produce soil on our own. In previous section, a sediment from pond (denoted as mud in the text) is treated by PVA and the PVA-mud exhibits high porosity, moderate wetness, and high drainage (Table 1), exhibiting improve property comparing to the original mud, which has low porosity and poor drainage (fig. s10, fig. s12 and fig. s14). This PVA-mud serves as an example for eroded soil renewal. The majority of eroded soil sediment in river and reservoir, creating sorts of environmental problem and being treated as solid waste. Apply our PVA treatment, these eroded soils can be returned to useful soil. Reduce erosion can only delay the depletion of soil, but to stop or to revert the tendency, a rapid production of soil is more urgent. Our PVA treatment method can suppress erosion and promote production simultaneously, it may be the solution to the crisis of soil depletion.

## References and Notes

1. R. Amundson *et al.*, Soil and human security in the 21st century. *Science* **348**, 1261071 (2015).
2. D. H. Wall, J. Six. (American Association for the Advancement of Science, 2015), vol. 347, pp. 695-695.
3. A. Koch *et al.*, Soil security: solving the global soil crisis. *Global Policy* **4**, 434-441 (2013).
4. IPCC, IPCC special report on climate change, desertification, land degradation, sustainable land management, food security, and greenhouse gas fluxes in terrestrial ecosystems. *IPCC Summary for Policymalers*, 1-472 (2019).
5. J. Kaiser. (American Association for the Advancement of Science, 2004).
6. P. Borrelli *et al.*, Land use and climate change impacts on global soil erosion by water (2015-2070). *Proceedings of the National Academy of Sciences* **117**, 21994-22001 (2020).
7. D. Pimentel *et al.*, Environmental and economic costs of soil erosion and conservation benefits. *Science* **267**, 1117-1123 (1995).

8. I. Shainberg, G. Levy, P. Rengasamy, H. Frenkel, Aggregate stability and seal formation as affected by drops' impact energy and soil amendments. *Soil Science* **154**, 113-119 (1992).
9. P. Kinnell, Raindrop‐impact‐induced erosion processes and prediction: a review. *Hydrological Processes: An International Journal* **19**, 2815-2844 (2005).
10. M. A. Nearing, S.-q. Yin, P. Borrelli, V. O. Polyakov, Rainfall erosivity: An historical review. *Catena* **157**, 357-362 (2017).
11. FAO, ITPS, Status of the world's soil resources (Main Report). (（FAO, 2015)).
12. W. Sun, Q. Shao, J. Liu, J. Zhai, Assessing the effects of land use and topography on soil erosion on the Loess Plateau in China. *Catena* **121**, 151-163 (2014).
13. P. Borrelli *et al.*, An assessment of the global impact of 21st century land use change on soil erosion. *Nature communications* **8**, 2013 (2017).
14. D. Hillel, *Introduction to soil physics*. (Academic press, 2013).
15. B. Barthès, E. Roose, Aggregate stability as an indicator of soil susceptibility to runoff and erosion; validation at several levels. *Catena* **47**, 133-149 (2002).
16. P. Farres, The dynamics of rainsplash erosion and the role of soil aggregate stability. *Catena* **14**, 119-130 (1987).
17. J. Liu *et al.*, Fatigue-resistant adhesion of hydrogels. *Nature communications* **11**, 1071 (2020).
18. E. Chiellini, A. Corti, B. Politi, R. Solaro, Adsorption/desorption of polyvinyl alcohol on solid substrates and relevant biodegradation. *Journal of Polymers and the Environment* **8**, 67-79 (2000).
19. M. Sjöberg, L. Bergström, A. Larsson, E. Sjöström, The effect of polymer and surfactant adsorption on the colloidal stability and rheology of kaolin dispersions. *Colloids and Surfaces A: Physicochemical and Engineering Aspects* **159**, 197-208 (1999).
20. L. Chen, S. H. Imam, S. H. Gordon, R. V. Greene, Starch-polyvinyl alcohol crosslinked film—performance and biodegradation. *Journal of environmental polymer degradation* **5**, 111-117 (1997).
21. E. Chiellini, A. Corti, S. D'Antone, R. Solaro, Biodegradation of poly (vinyl alcohol) based materials. *Progress in Polymer science* **28**, 963-1014 (2003).
22. L. R. Krupp, W. J. Jewell, Biodegradability of modified plastic films in controlled biological environments. *Environmental science & technology* **26**, 193-198 (1992).
23. H. Sawada, in *Studies in polymer science*. (Elsevier, 1994), vol. 12, pp. 298-312.
24. M. Kimura, K. Toyota, M. Iwatsuki, H. Sawada, in *Studies in Polymer Science*. (Elsevier, 1994), vol. 12, pp. 92-106.
25. C. DeMerlis, D. Schoneker, Review of the oral toxicity of polyvinyl alcohol (PVA). *Food and chemical Toxicology* **41**, 319-326 (2003).

**Acknowledgments:** The manuscript is drafted by the author, and ChatGPT-3.5 is used to check and revise grammar errors.

**Funding:**

Young Innovative Talents Project of Universities in Guangdong Province, China 2022KQNCX137 (CC)

**Author contributions:**

Conceptualization: GL

Methodology: GL

Funding acquisition: CC

Project administration: CC, GL

Writing – original draft: GL

Writing – review & editing: CC, GL

**Competing interests:** The authors declare that they have no conflicts of interest. They also affirm that they will not apply any patents related to this paper and assure that their technique is freely accessible to all..

**Data and materials availability:** All data are available in the main text or the supplementary materials.

**Supplementary Materials**

Materials and Methods

Figs. S1 to S16

Movies S1 to S4

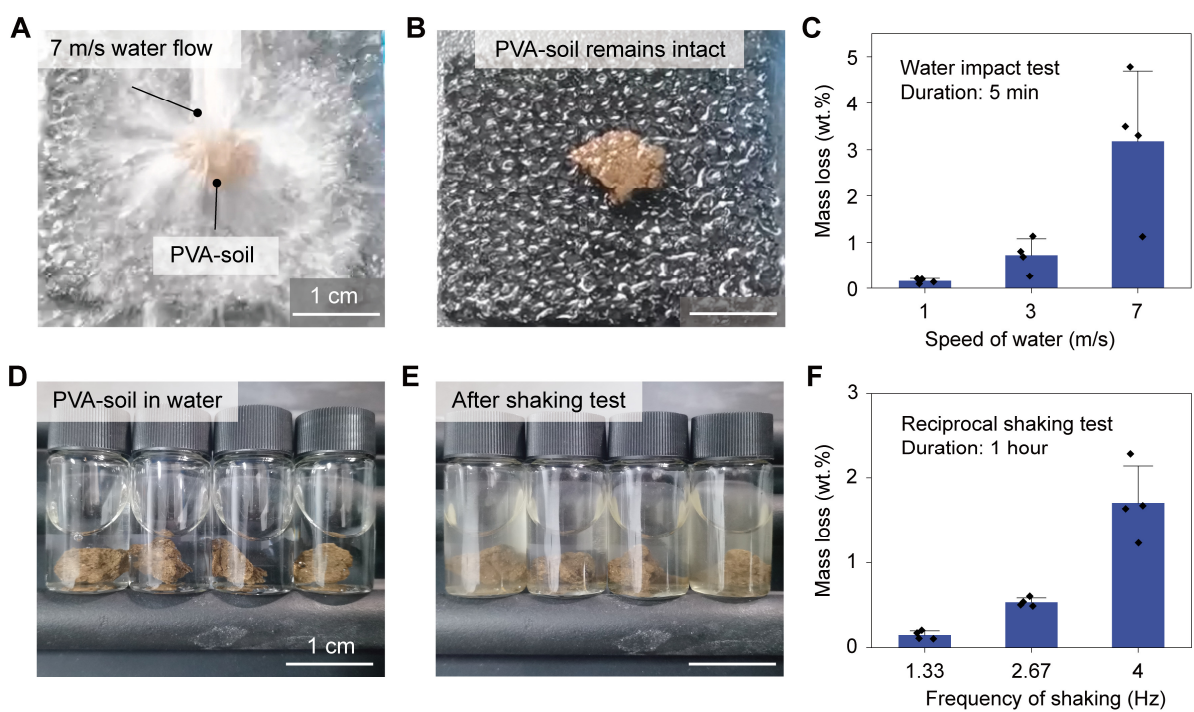

**Fig. 1. Stability of PVA-soil in water.** (A). PVA-soil under the impact of 7 m/s water reveals no fracture. (B). PVA-soil remains intact after the water impact test. (C). Mass loss of PVA-soil in water impact test at different flowing speed, the test duration is 5 min. (D). PVA-soil in vial before the shaking test. (E). PVA-soil in vial after 1 hour of shaking at 4 Hz, PVA-soil maintains intact, only small fraction of soil is disintegrated and making the water slightly cloudy. (F). Mass loss of PVA-soil in reciprocal shaking test at different shaking frequencies, shaking duration is 1 hour. Error bar represents standard deviation.

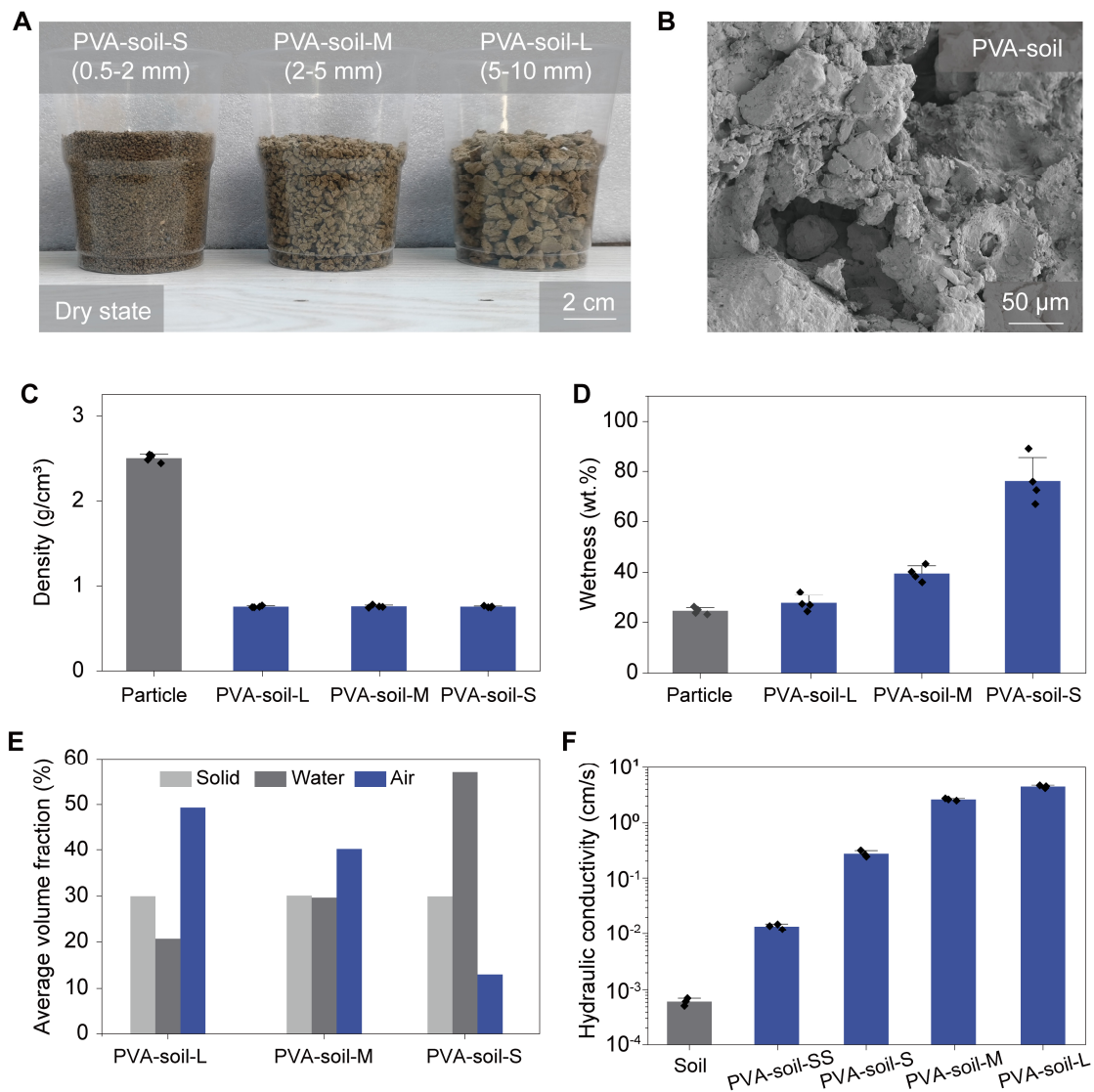

**Fig. 2. Porous structure and physical properties of PVA-soils at different sizes.** (A). Photo of three PVA-soils. The size of PVA-soil-S ranges from 0.2 to 2 mm, indicating that the sample passes through the sieve with a mesh size of 2 mm but is retained by the sieve with a mesh size of 0.2 mm. The same denotation applies to PVA-soil-M and PVA-soil-L. (B). Internal microstructure of PVA-soil captured by scan electron microscopy reveals large pore in dimension of 100 μm and small pore in several micrometer. (C). Density of PVA-soils. The particle density is measured by pycnometer method. The bulk density of PVA-soil-S/M/L is measured by packing the sample into a container and then dividing the mass by the volume. (D). Wetness of PVA-soils. The wetness of particle is measured in the form of single grain. The wetness of PVA-soil-S/M/L is measured when the sample is packed in a container. (E). Volume fractions of solid, liquid and gas of PVA-soils at saturated state. (F). Hydraulic conductivity of PVA-soils. The hydraulic conductivity of untreated soil is measured by a variable head method while the PVA-soils are measured by a constant head method. Error bar represents standard deviation.

**Table 1. Physical properties of PVA treated soils.**

| Items[a] | Mass loss (wt.%) | Bulk density (g cm$^{-3}$) | Porosity (%) | Wetness (wt.%) | Air fraction (vol.%) | Hydraulic conductivity (cm s$^{-1}$) |
|---|---|---|---|---|---|---|
| PVA-soil-L | 3.17±1.52 | 0.75±0.01 | 70.0±0.4 | 27.6±3.3 | 49.3 | 4.50±0.21 |
| PVA-soil-ver | 3.65±1.04 | 0.65±0.02 | 73.4±0.6 | 37.5±1.1 | 49 | 5.22±0.10 |
| PVA-soil-per | 2.72±1.03 | 0.56±0.01 | 78.4±0.3 | 35.1±1.8 | 58.8 | 6.12±0.26 |
| PVA-mud | 4.22±0.80 | 0.69±0.02 | 71.7±0.7 | 28.6±0.8 | 51.9 | 4.64±0.16 |
| PVA-red soil | 6.33±0.87 | 0.81±0.02 | 67.1±0.6 | 26.0±0.9 | 48.8 | 4.32±0.16 |

[a] The size of all five samples falls within the range of 5-10 mm. The same testing method is applied to each sample to obtain the respective properties.

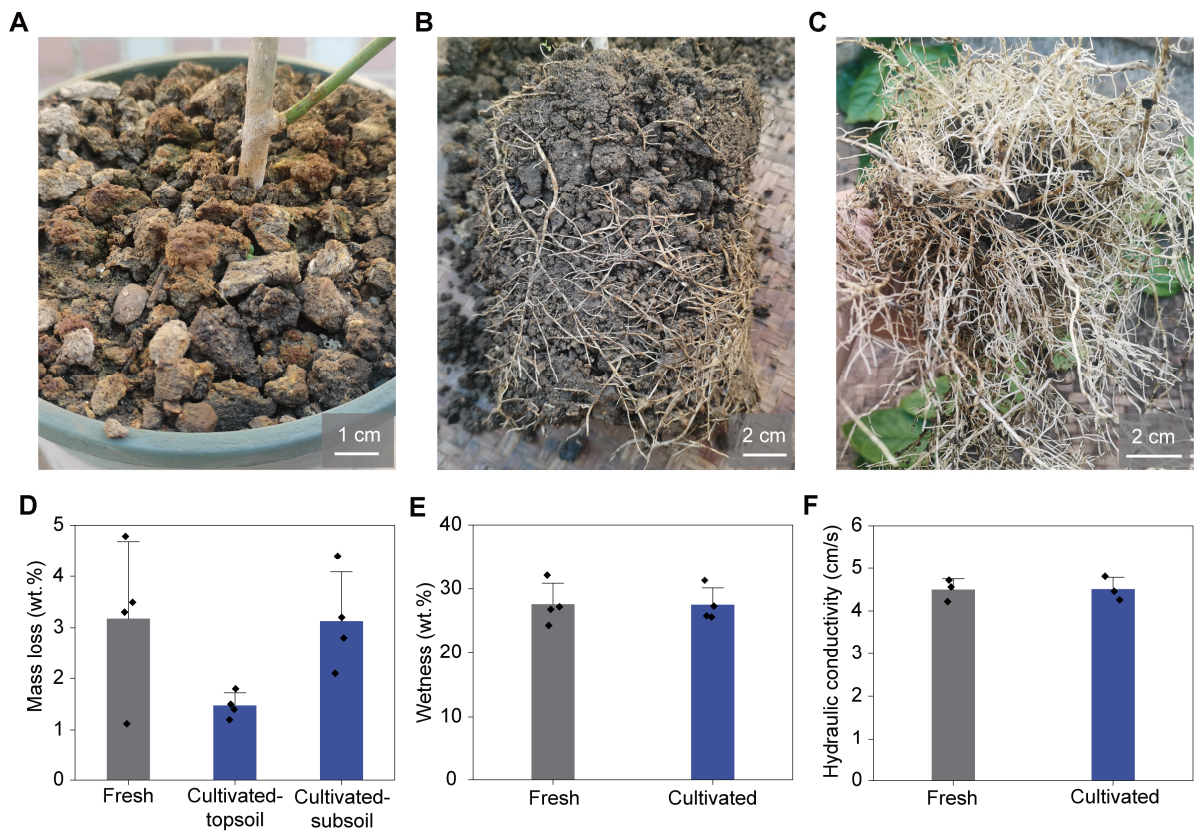

**Fig. 3. Porous structure and physical properties of PVA-soils at different sizes.** (A). Surface PVA-soil after eight months cultivation reveals no decomposition with growth of lichen and moss. (b). Subsoil of the pot shows discrete PVA-soils grains and leaving large pores. (c). Root of the cultivated plant shows a dense and uniform growth. (d). Mass loss of the fresh and cultivated PVA-soil in water impact test at 7 m/s, impact duration is 5 min. (e). Wetness of the fresh and cultivated PVA-soil. The size of these two soils is the same, ranging from 5-10 mm. (F). Drainage of the fresh and cultivated PVA-soil. The soils are in the same size of 5-10 mm. Error bar represents standard deviation.

# Supplementary Materials for

## Soil Stabilization and Renewal Using Annealed Polyvinyl Alcohol as Long-lasting Binder

Chunyan Cao[1]* and Gang Li[2]*

Corresponding author: caochy@nfu.edu.cn and 11930690@mail.sustech.edu.cn

**The PDF file includes:**

    Materials and Methods
    Figs. S1 to S16

**Other Supplementary Materials for this manuscript include the following:**

    Movies S1 to S4

**Materials and Methods**

Preparation of PVA solution
Three grams of PVA powder (KURARAY POVAL™ 60-98, 98-99 mol% hydrolysis) is added to 97 g of water and stirred in a 95 °C water bath until completely dissolved.

Preparation of PVA-soil
One hundred grams of dried soil is mixed with 67 g of PVA solution. The mixture is dried in a 60 °C oven for 1 day and annealed at 120 °C for 4 hours. The PVA-soil is then crushed into small particles. The crushed PVA-soil is sieved into different particle sizes.

Preparation of PVA hydrogel
Ten grams of the PVA solution are cast onto a PET plate and dried at room temperature for 1 day, followed by annealing at 120 °C for 4 hours.

Preparation of clay-PVA composite
One hundred grams of kaolin or montmorillonite is mixed with 100 g of PVA solution. The mixtures are dried in a 60 °C oven for 1 day, followed by annealing at 120 °C for 4 hours. The clay-PVA composite is then crushed and sieved.

Wet sieving test
Dried untreated soil and PVA-soil are crushed and sieved into size ranging from 5-10 mm. These soil aggregates are placed above the sieve of mesh size of 5 mm. Then the sieve is immersed in water bath and moving up and down with a displacement of 3 cm and a frequency of 30 time per minute. Then the sample leaving on the sieve is collected and fully dried again. The change in mass before and after this sieving process is used to indicate the water stability of the soils.

Water impact test
PVA-soil is pre-soaked in water bath for 1 day before the test. The saturated PVA-soil is weighed and placed under the direct impact of water for 5 min at 1 m/s, 3 m/s or 7 m/s. Then the PVA-soil is weighed again and the reduction in mass is recorded.

Reciprocal shaking test
PVA-soil is pre-soaked in water bath for 1 day before the test. The saturated PVA-soil is weighed and placed into a 20-mL vial containing about 15 g water. The vial is fixed on the reciprocal shaker and shaking at set frequencies and time. After that the PVA-soil is removed from the vial and weighed again. The reduction in mass is used to measure stability of PVA-soil in shaking test.

Measurement of particle density
Ten gram of dried PVA-soil is placed into a 50-mL pycnometer, then 20 g water is added into the bottle. The bottle is kept in vacuum for 1 hour to remove bubbles. Finally the bottle is transfer to atmosphere and filled with water. The mass of the bottle, soil and water is weighed as $m_{bws}$. The same pycnometer only filled with water is weighed as $m_{bw}$. The particle density of PVA-soil, $\rho_P$ is determined as:

$$\rho_P = \frac{m_s}{m_s + m_{bw} - m_{bws}} \times \rho_w$$

Where $m_s$ is the mass of PVA-soil and $\rho_w$ is the density of water at 25 °C (since the experiment is conducted at room temperature).

Measurement of bulk density and porosity
The sieved PVA-soil are filled into a plastic collum with volume of 1 L. Oscillation and percussion is applied to enhance package. Then the PVA-soil is weighed and the bulk density $\rho_b$ is calculated by dividing mass to volume.
The porosity (*f*) of PVA-soil is determined as:
$$f = 1 - \frac{\rho_b}{\rho_p}$$

Measurement of wetness
One kilogram of PVA-soil is packed in a plastic pot and filled with water. Then the pot is transferred in a vacuum environment for 1 hour to remove air bubbles. After that the pot is took out and four holes of about 3 mm is drilled in the bottom of the pot to drain the water. A cover is placed on top to reduce water evaporation. After 12 hours, the drained pot is weighed and the increase in mass is associated to the absorbed water and used to calculate the wetness.

Measurement of hydraulic conductivity
The PVA-treated soil was placed in a cylindrical apparatus with a cross-sectional area of 30 cm². Under constant head conditions and a steady vertical water flow, the volume of water passing through the specimen was measured over a specific time interval. Subsequently, the hydraulic conductivity of the specimen was calculated using Darcy's law.
For untreated soil, hydraulic conductivity is measured by a falling head condition.

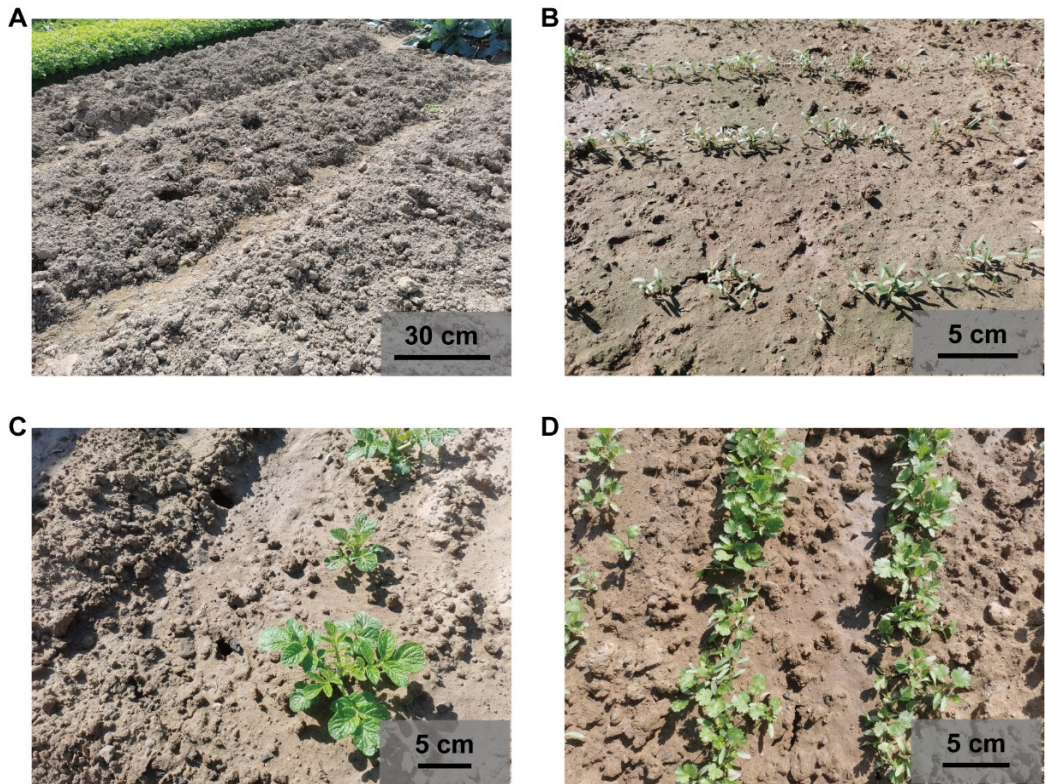

**Fig. S1.**
**Soil at the collection site.**
Bare soil (A) and soil planted with cilantro (B), potato (C), and celery (D). The collection site is in Chaozhou, Guangdong, China. The soil has been used for vegetable cultivation for over 10 years, and the overuse of fertilizer has led to soil degradation, as indicated by the formation of crust observed in areas B, C, and D.

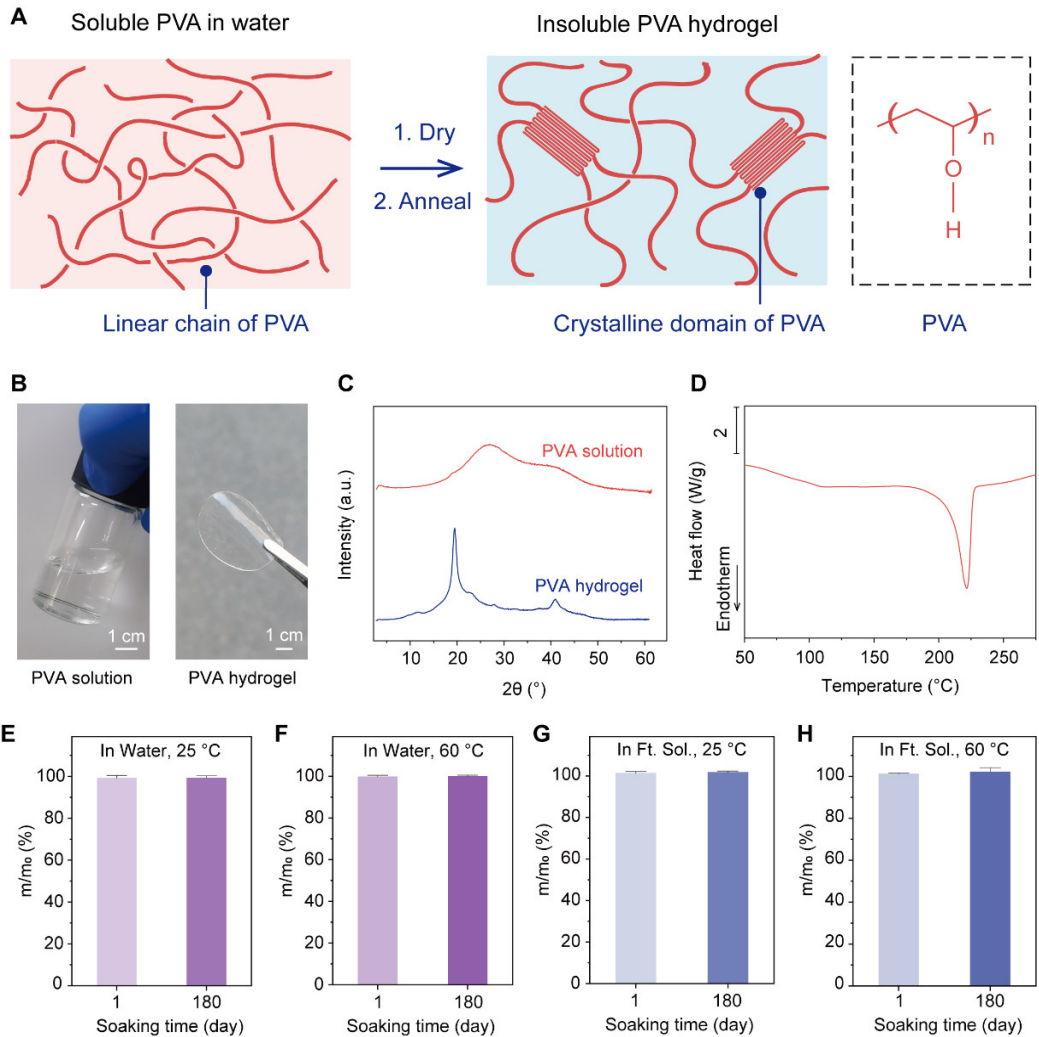

**Fig. S2.**
**Properties of annealed PVA.**
(A) Drying and annealing the PVA aqueous solution remove the water and induce the crystallization of PVA, resulting in a three-dimensional polymer network crosslinked by crystalline domains of PVA. (B) Photo of PVA solution and PVA hydrogel. (C) Wide-angle X-ray scattering profiles reveal a newly formed diffraction peak around 2θ=19° in the hydrogel formed by annealed PVA. This peak corresponds to the crystal surface of PVA, indicating the formation of a crystalline domain. (D) The annealed PVA shows an endothermic peak around 220 °C, which originates from the melting of the crystalline domain of PVA. Drying and annealing treatment endow PVA with high stability in water and in a fertilizer solution. No mass loss is observed in the PVA hydrogel after soaking in water and a fertilizer solution at room temperature and at 60 °C for 180 days. (E-H). The high stability of the PVA hydrogel prevents it from being carried away by water. In comparison, other water-soluble polymer conditioners such as PAM suffer from the problem of a short working duration. The major reason is that soluble polymers can be washed away by water, such as rainfall or irrigation. The drying and annealing treatment reverts the PVA into water-insoluble form, thus preventing the wash-away problem.

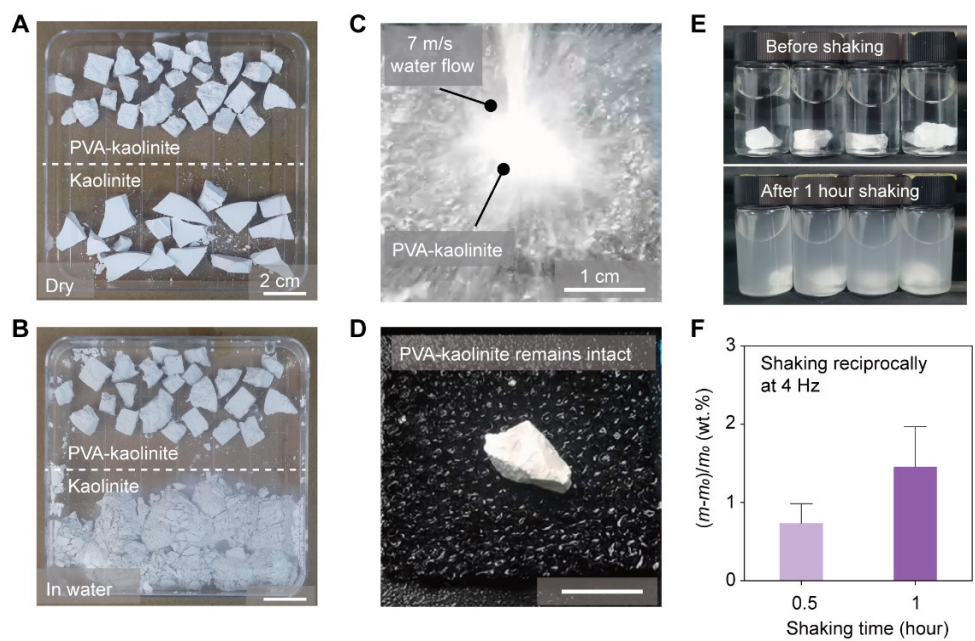

**Fig. S3.**
**Properties of PVA-kaolinite composite.**
Sand, silt, and clay are the mineral components of soils, where sand and silt are silica, and clay is layered aluminosilicate. Clay can absorb PVA in an irreversible manner, providing the possibility to bind clay using PVA. Kaolinite is selected as the representative example of clay and is bound with 3 wt.% PVA through mixing, drying, and annealing. The PVA-kaolinite composite does not disintegrate in water, whereas kaolinite collapses in the water bath(A, B). The PVA-kaolinite composite can withstand high-speed water impact. A water flow of 7 m/s does not break the PVA-kaolinite (C, D). The PVA-kaolinite can sustain reciprocally shaking at 4 Hz for 1 hour with a mass loss of 2% (E, F). Such high stability of PVA-kaolinite is ascribed to the strong adhesion between PVA and kaolinite. The long-chain PVA cements the discrete clay particles into a form like a necklace, where the PVA forms a 3D thread network and the kaolinite is the bead on the thread. Without PVA, removing one kaolinite particle from a grain of kaolinite only requires overcoming the weak interaction between adjacent clay particles. The insertion of water molecules causes the adjacent clays to separate. In contrast, if one wants to separate a kaolinite particle from the PVA-kaolinite composite, it is necessary to overcome the interaction between kaolinite and its absorbed PVA, or cut all the absorbed PVA chains. Both processes require a high input of energy that surpasses the kinetic energy from high-speed water impact and high-frequency shaking. The mixing-drying-annealing treatment is vital to bestow high adhesion between PVA and clay. The mixing step ensures the uniform distribution of PVA among the kaolinite composite, leading to the initial absorption of PVA at the active positions of kaolinite. The drying step removes a large amount of water to enhance the contact between PVA and kaolinite. The annealing treatment eliminates residual water molecules and ensures that the contact between PVA and kaolinite reaches the highest level. Additionally, the annealing treatment induces crystallization in unabsorbed PVA and increases its stability in water (Fig. S1). By binding the clay particles of the soil, PVA can improve soil stability. Although PVA cannot bind sand and silt, the network created by PVA and clay is capable of trapping sand and silt inside.

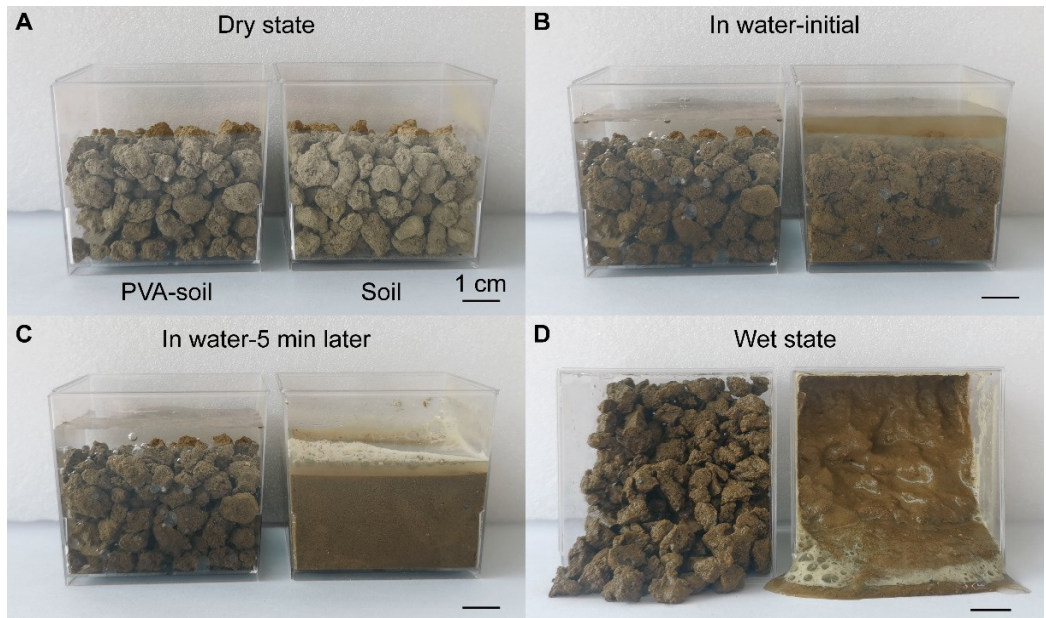

**Fig. S4.**
**Appearance of PVA-soil and untreated soil in the dry state and the wet state.**
(A) Photo of PVA-soil and untreated soil in the dry state, with their sizes falling within the same range of 5-10 mm. (B) Water does not break the PVA-soil, but it causes the untreated soil to disintegrate. (C) PVA-soil does not change after soaking for 5 minutes, while the untreated soil disintegrates into smaller particles. (D) The PVA-soil absorbs water and remaining initial shape, while the untreated soil turns into a mud-like consistency.

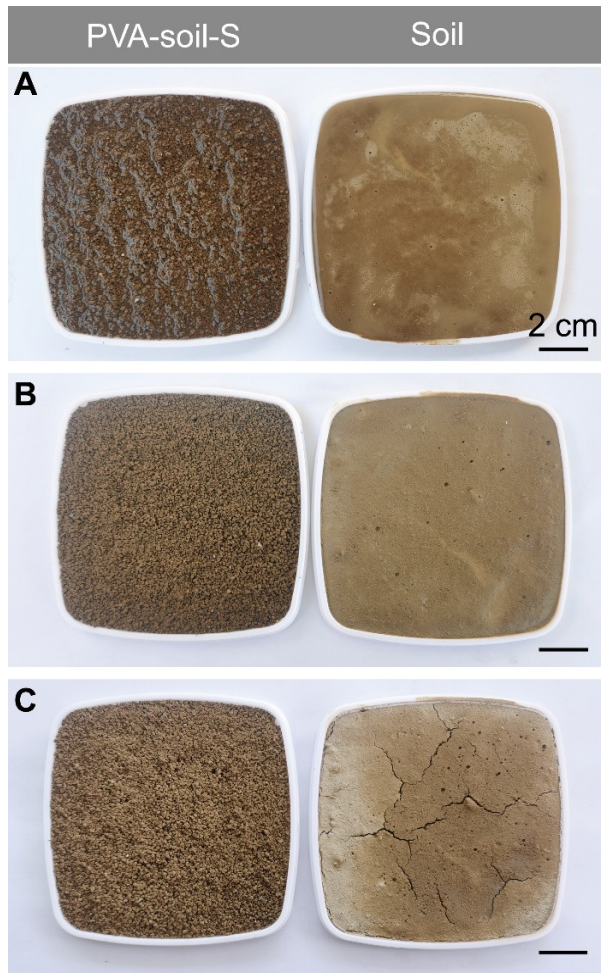

**Fig. S5.**
**Appearance of PVA-soil and untreated soil during the drying process.**
The PVA-soil, with a size of 0.5-2 mm, maintains its discrete particle form during the drying process, indicating its inertness to water (A, B, and C). In contrast, the untreated soil transforms into a continuous film when partially dried (A, B). Further drying causes the soil to crack (C).

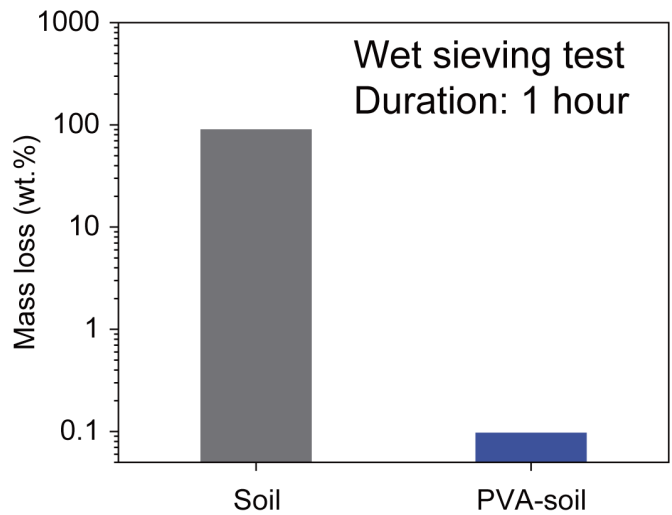

**Fig. S6.**
**Mass changes of soil and PVA-soil in a wet sieving test.**
The untreated soil and PVA-soil, both ranging in size from 5-10 mm, are placed on a sieve with a mesh size of 5 mm and subjected to wet sieving. The sample that remains on the sieve after the tests are collected and weighed. This weight is then compared with its initial mass to calculate the mass loss.

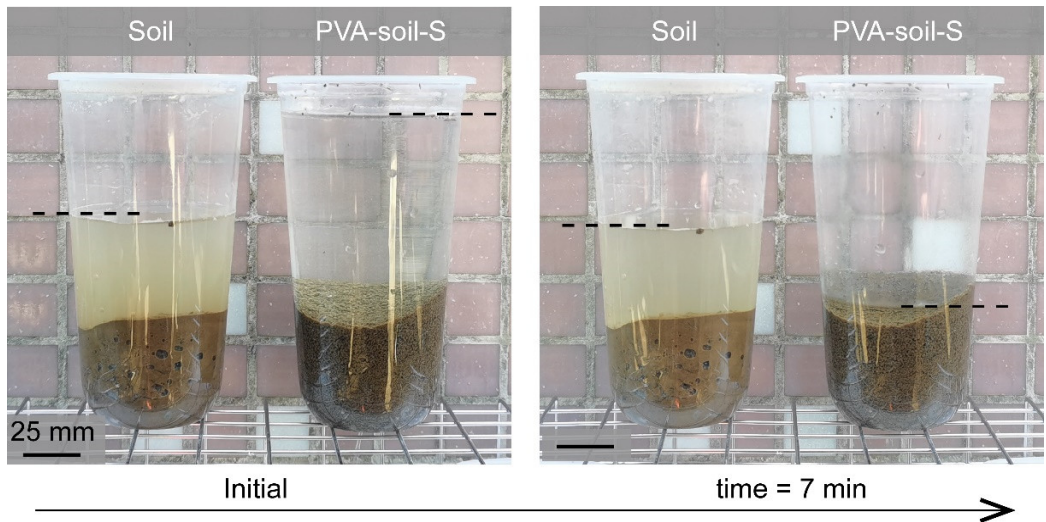

**Fig. S7.**
**Demonstration of drainage properties in soil and PVA-soil-S.**
Both of the untreated soil and PVA-soil-S are placed in a cup, which has a 2 mm hole in the bottom for water drainage. A larger amount of water is added to the cup containing PVA-soil-S initially. After 7 minutes, the water in the cup with PVA-soil-S is mostly drained, whereas the cup with soil shows only a minor drop in water level.

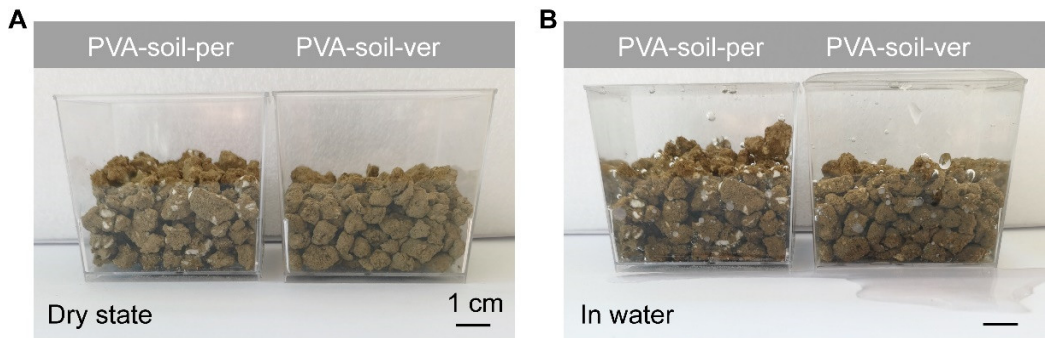

**Fig. S8.**
**PVA-soil-per and PVA-soil-ver in dry state and in water.**
The addition of perlite or vermiculite does not deteriorate the stability of the PVA-treated sample, revealing no disintegration in water (A, B).

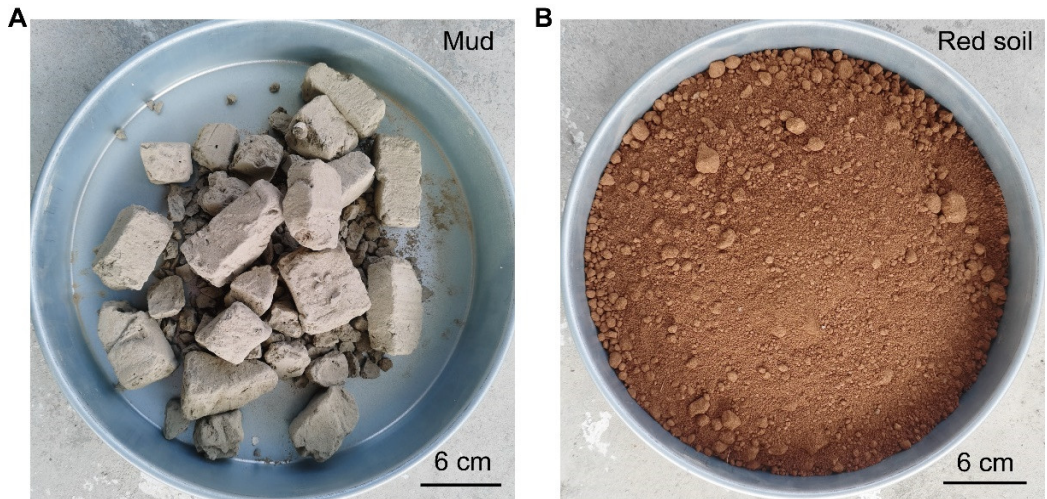

**Fig. S9.**
**Photo of the received mud and red soil.**
(A) Mud is the dried sediment from a pond in Chaozhou. (B) Red soil is from Longyan, Fujian, China.

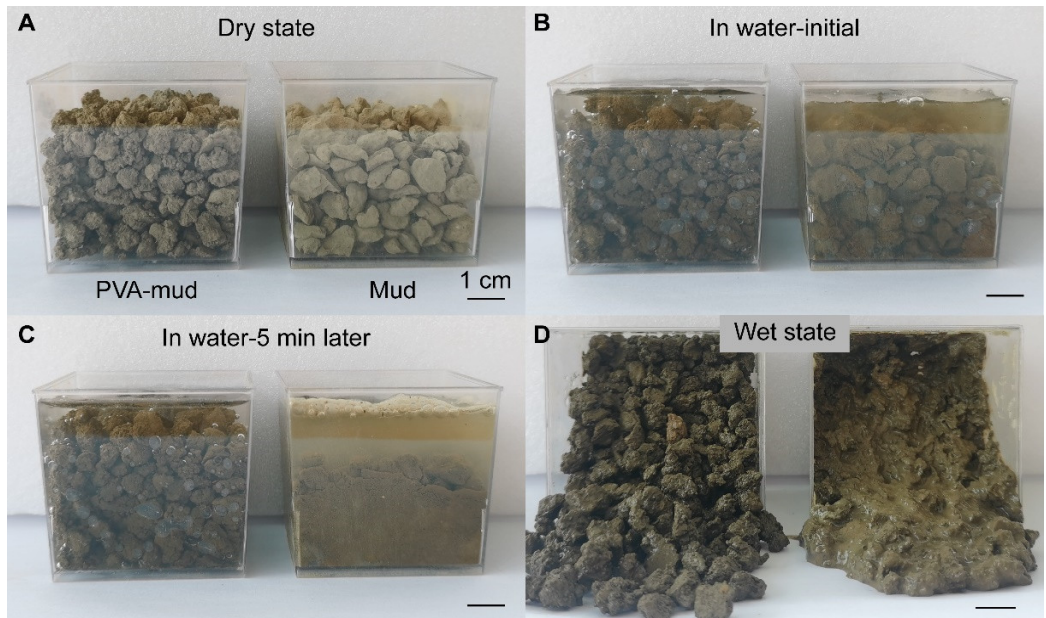

**Fig. S10.**
**Appearance of PVA-mud and untreated mud in the dry state and the wet state.**
(A) Photo of PVA-mud and untreated mud in the dry state, with their sizes falling within the same range of 5-10 mm. (B) Water does not break the PVA-mud, but it causes the untreated mud to disintegrate. (C) PVA-mud does not change after soaking for 5 minutes, while the untreated mud disintegrates into smaller particles. (D) PVA-mud absorbs water and retains its initial shape, while the untreated mud loses its initial shape.

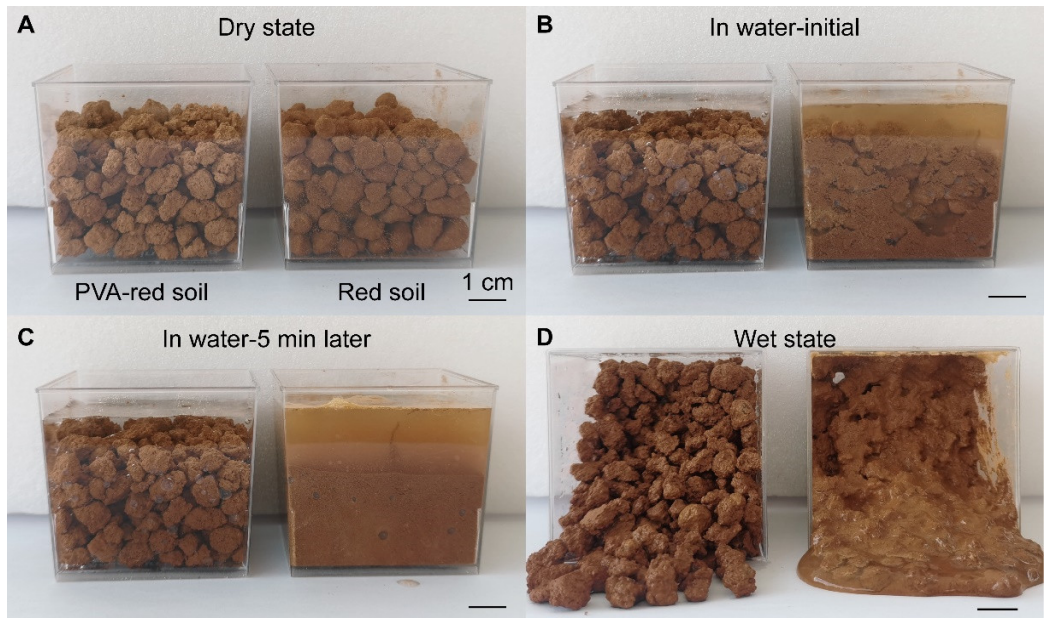

**Fig. S11.**
**Appearance of PVA-red soil and untreated red soil in the dry state and the wet state.**
(A) Photo of PVA-red soil and untreated red soil in the dry state, with their sizes falling within the same range of 5-10 mm. (B) Water does not break the PVA-red soil, but it causes the untreated red soil to disintegrate. (C) PVA-red soil does not change after soaking for 5 minutes, while the untreated red soil disintegrates into smaller particles. (D) PVA-red soil absorbs water and retains its initial shape, while the untreated red soil loses its initial shape.

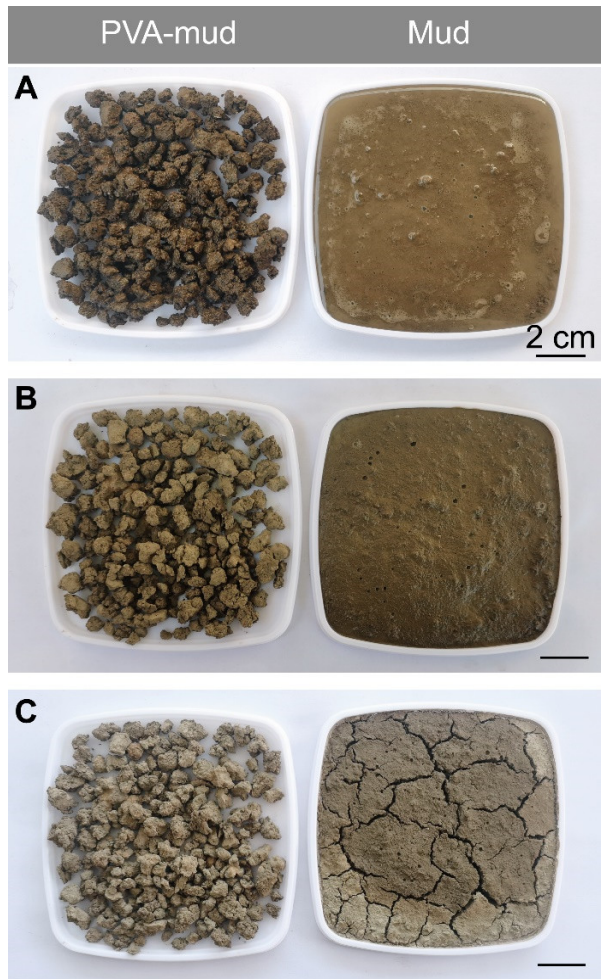

**Fig. S12.**
**Appearance of PVA-mud and untreated mud during the drying process.**
The PVA-mud, with a size of 5-10 mm, maintains its discrete particle form during the drying process, indicating its inertness to water (A, B, and C). In contrast, the untreated mud transforms into a continuous film when partially dried (A, B). Further drying causes the mud to crack (C).

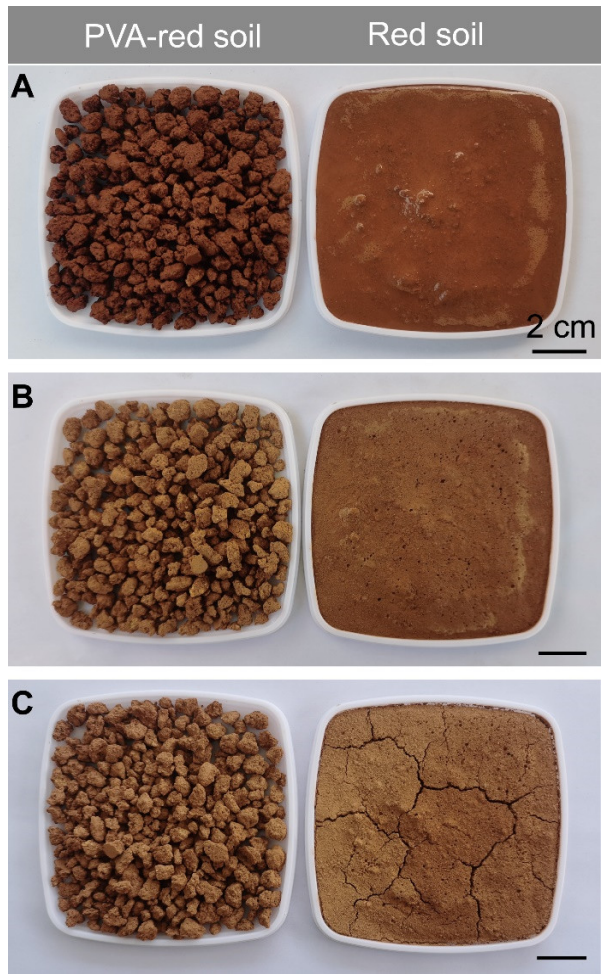

**Fig. S13.**
**Appearance of PVA-red soil and untreated red soil during the drying process.**
The PVA-red soil, with a size of 0.5-2 mm, maintains its discrete particle form during the drying process, indicating its inertness to water (A, B, and C). In contrast, the untreated red soil transforms into a continuous film when partially dried (A, B). Further drying causes the red soil to crack (C).

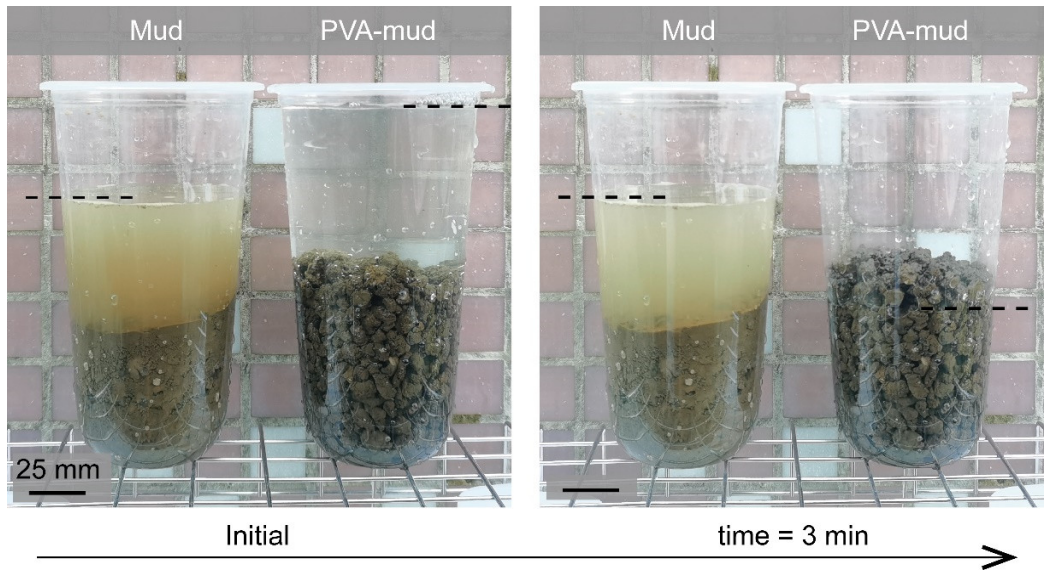

**Fig. S14.**
**Demonstration of drainage properties in mud and PVA-mud.**
Both of the untreated mud and PVA-mud are placed in a cup, which has a 2 mm hole in the bottom for water drainage. A larger amount of water is added to the cup containing PVA-mud initially. After 3 minutes, the water in the cup with PVA-mud is mostly drained, whereas the cup with mud shows only a minor drop in water level.

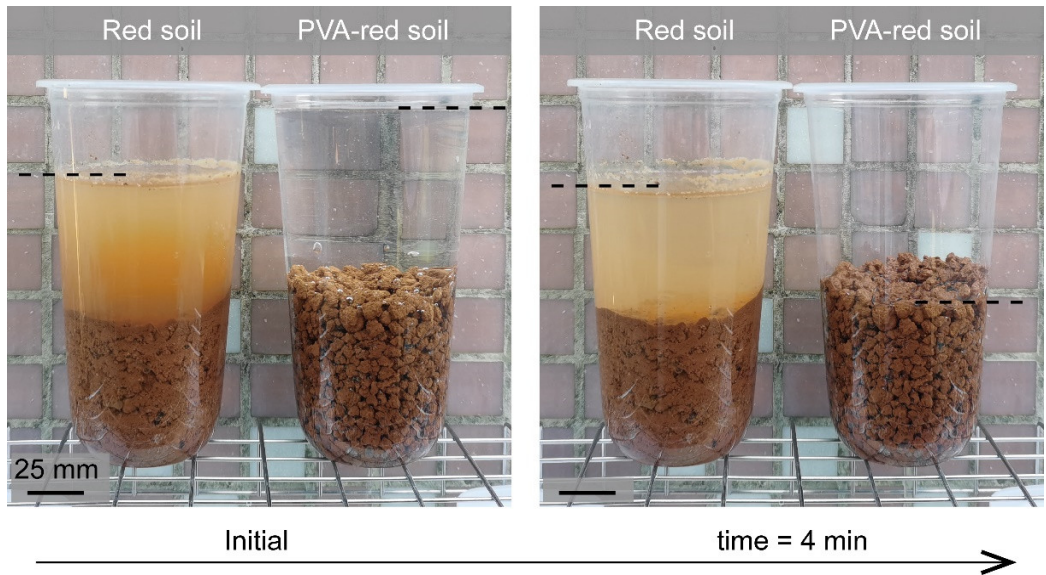

**Fig. S15.**
**Demonstration of drainage properties in red soil and PVA-red soil.**
Both of the untreated red soil and PVA-red soil are placed in a cup, which has a 2 mm hole in the bottom for water drainage. A larger amount of water is added to the cup containing PVA-red soil initially. After 4 minutes, the water in the cup with PVA-red soil is mostly drained, whereas the cup with red soil shows only a minor drop in water level.

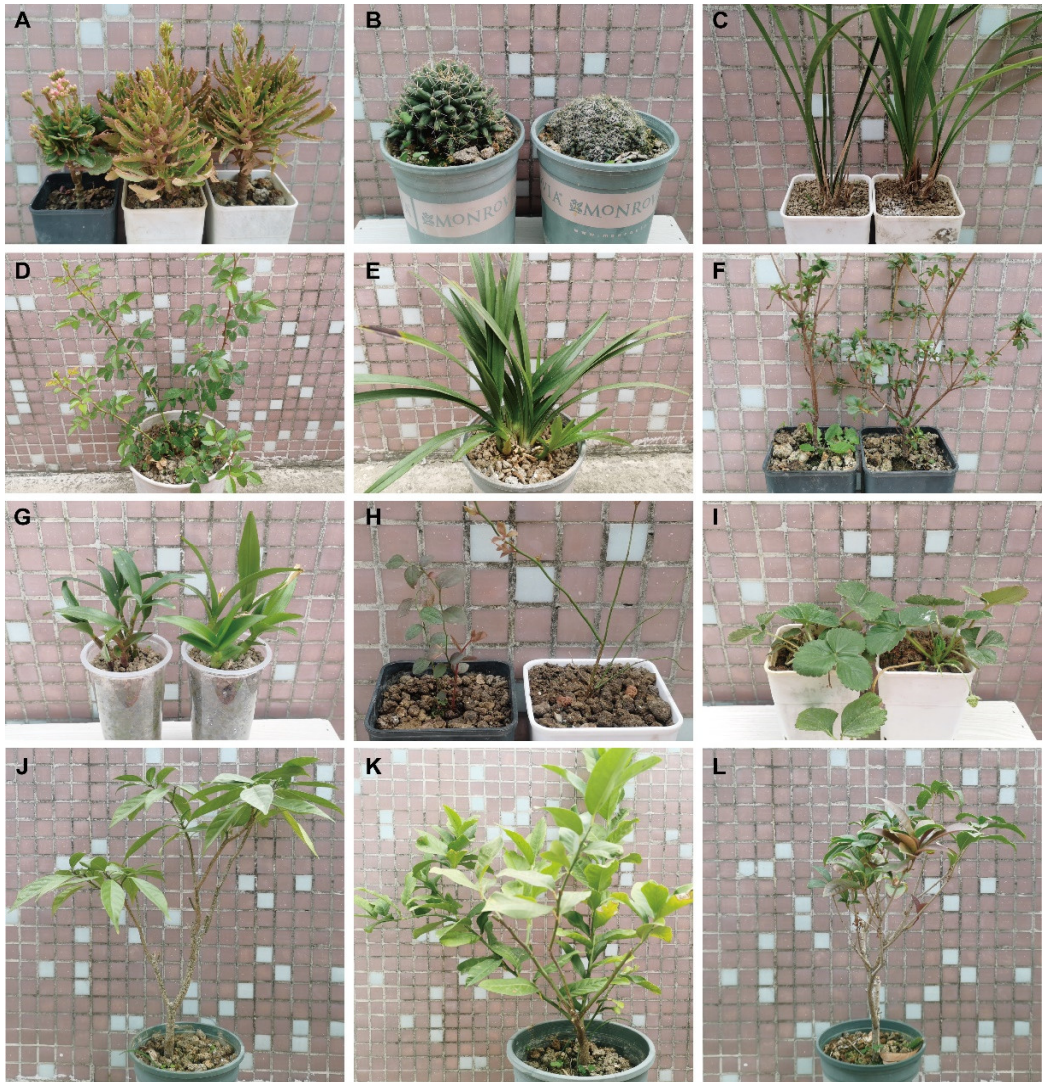

**Fig. S16.**
**Photo of plants cultivated using PVA-soil as substrate.**
The species of plants are listed as follows: (A) Kalanchoe blossfeldiana. (B) Cactus. (C) Orchid. (D) rose. (E) Cymbidium sp.. (F) Rhododendron indicum. (G) Dendrobium. (H) Cyanococcus. (I) . Strawberry (J) Ervatamia. (K) Lemon. (L) Osmanthus fragrans. These plants belong to multiple genera and demonstrate robust growth, showcasing the excellent performance of PVA-soil as a planting substrate. There is no scale bar in the image; however, the mosaic bricks in the background serve as a scale, with the length of each square bricks measuring 2.5 cm.

**Movie S1.**
Water impact test on PVA-soil.

**Movie S2.**
Reciprocal shaking test on PVA-soil.

**Movie S3.**
Drainage of soil and PVA-soil.

**Movie S4.**
Drainage of cultivated soil and culticated PVA-soil.